\documentclass[aps,pra,twocolumn,longbibliography]{revtex4-1}
\usepackage{amsmath}
\usepackage{epsfig}
\usepackage{epstopdf}
\usepackage{graphicx}
\graphicspath{{figures/}}
\usepackage{rotating}
\usepackage{amssymb}
\usepackage{float}
\usepackage[
final,inline,nomargin]{fixme} \fxsetup{theme=color}
\usepackage{braket}
\usepackage{subfigure}
\usepackage{natbib}
\usepackage{color}

\fxsetup{status=draft}

\newcommand{\1}{{\text A}}
\newcommand{\2}{{\text B}}
\newcommand{\evec}{\ensuremath{\boldsymbol{\eta}}}

\begin{document}

\title{Few-body bound states of two-dimensional bosons}

\author{G. Guijarro}
\author{G. E. Astrakharchik}
\author{J. Boronat}
\affiliation{Departament de F\'isica, Campus Nord B4-B5, Universitat Polit\`ecnica de Catalunya, E-08034 Barcelona, Spain}
\author{B. Bazak}
\affiliation{The Racah Institute of Physics, The Hebrew University, 9190401, Jerusalem, Israel}
\author{D. S. Petrov}
\affiliation{Universit\'e Paris-Saclay, CNRS, LPTMS, 91405, Orsay, France}

\date{7th of April, 2020}

\begin{abstract}
We study clusters of the type A$_N$B$_M$ with $N\leq M\leq 3$ in a two-dimensional mixture of A and B bosons, with attractive AB and equally repulsive AA and BB interactions. In order to check universal aspects of the problem, we choose two very different models: dipolar bosons in a bilayer geometry and particles interacting via separable Gaussian potentials. We find that all the considered clusters are bound and that their energies are universal functions of the scattering lengths $a_{\1\2}$ and $a_{\1\1}=a_{\2\2}$, for sufficiently large attraction-to-repulsion ratios $a_{\1\2}/a_{\2\2}$. When $a_{\1\2}/a_{\2\2}$ decreases below $\approx 10$, the dimer-dimer interaction changes from attractive to repulsive and the population-balanced AABB and AAABBB clusters break into AB dimers. Calculating the AAABBB hexamer energy just below this threshold, we find an effective three-dimer repulsion which may have important implications for the many-body problem, particularly for observing liquid and supersolid states of dipolar dimers in the bilayer geometry. The population-imbalanced ABB trimer, ABBB tetramer, and AABBB pentamer remain bound beyond the dimer-dimer threshold. In the dipolar model, they break up at $a_{\1\2}\approx 2 a_{\2\2}$ where the atom-dimer interaction switches to repulsion.
\end{abstract}

\maketitle

Recent experiments on dilute quantum droplets in dipolar bosonic gases~\cite{Kadau2016,Schmitt2016,Ferrier2016,Chomaz2016} and in Bose-Bose mixtures~\cite{Cabrera2017,Semeghini2018,Ferioli2019} with competing interactions have exposed the important role of beyond-mean-field effects in weakly-interacting systems. A natural strategy to boost these effects and enhance exotic behaviors is to make the interactions stronger while keeping the attraction-repulsion balance for mechanical stability. The most straightforward way of getting into this regime is to increase the gas parameter $na_s^3$. However, this leads to enhanced three-body losses which results in very short lifetimes (as it has been observed in experiments~\cite{Kadau2016,Schmitt2016,Ferrier2016,Chomaz2016,Cabrera2017,Semeghini2018,Ferioli2019}). Nevertheless, this regime is achievable in reduced geometries. It has been shown that a one-dimensional Bose-Bose mixture with strongly-attractive interspecies interaction becomes dimerized and, by increasing the intraspecies repulsion, the dimer-dimer interaction can be tuned from attractive to repulsive~\cite{Pricoupenko2018}. Then, an effective three-dimer repulsion has been found in this system and predicted to stabilize a liquid phase of attractive dimers~\cite{Guijarro2018}. 

In two dimensions, a particularly interesting realization of such a strongly-interacting, tunable, and long-lived Bose-Bose mixture is a system of dipolar bosons confined to a bilayer geometry
\cite{WangLukinDemler2006,Wang2007,Trefzger_2011}.
When the dipoles are oriented perpendicularly to the plane, there is a competing effect between repulsive intralayer and partially attractive interlayer interactions, interesting from the viewpoint of liquid formation. 
In addition, the quasi-long range character of the dipolar interaction can produce the rotonization of its spectrum and a supersolid behavior~\cite{ODell2003,Santos2003,Lu2015,Tanzi2019,Chomaz2019,Bottcher2019,Tanzi2019Modes,Guo2019,Ferlaino2019}, formation of a crystal phase~\cite{Buchler2007,Astrakharchik2007}, and a pair superfluid \cite{Macia2014,PhysRevA.94.063630,Nespolo_2017} (see also lattice calculations of Ref.~\cite{Safavi2017}). A peculiar feature of bilayer model is the vanishing Born integral for the interlayer interaction, $\int V_{\1\2}(\rho)d^2\rho = 0$ \cite{bitube}, which has led to controversial claims about the existence of a two-body bound state~\cite{Yudson1997} till it has finally been established that this bound state always exists, although its energy can be exponentially small~\cite{Shih2009,Armstrong2010,Klawunn2010,Baranov2011,Volosniev2011}, consistently with Ref.~\cite{Simon1976}. Interestingly, a similar controversy seems to continue at the few-body level; it has been claimed~\cite{Volosniev2012} that the repulsive dipolar tails will never allow for three- or four-body bound states in this geometry.

In this Rapid Communication, we investigate few-body bound states in a two-dimensional mass-balanced mixture of A and B bosons with two types of interactions characterized by the two-dimensional scattering lengths $a_{\rm AB}$ and $a_{\rm AA}=a_{\rm BB}$. The first case corresponds to the bilayer of dipoles discussed above and, in the second, we model the interactions by non-local (separable) finite-range Gaussian potentials. By using the diffusion Monte Carlo (DMC) technique in the first case, and the Stochastic Variational Method (SVM) in the second, we find that for sufficiently weak BB repulsion compared to the AB attraction, $a_{\rm AB}\gg a_{\rm BB}$, all clusters of the type A$_N$B$_M$ with $1 \leq N \leq M \leq 3$ are bound. We then locate thresholds for their unbinding with decreasing $a_{\1\2}/a_{\2\2}$. By looking at the AAABBB hexamer energy close to the corresponding threshold, we discover an effective three-dimer repulsion, which can stabilize interesting many-body phases.

\begin{figure}[H]
	\centering
	\includegraphics[width=0.45\textwidth]{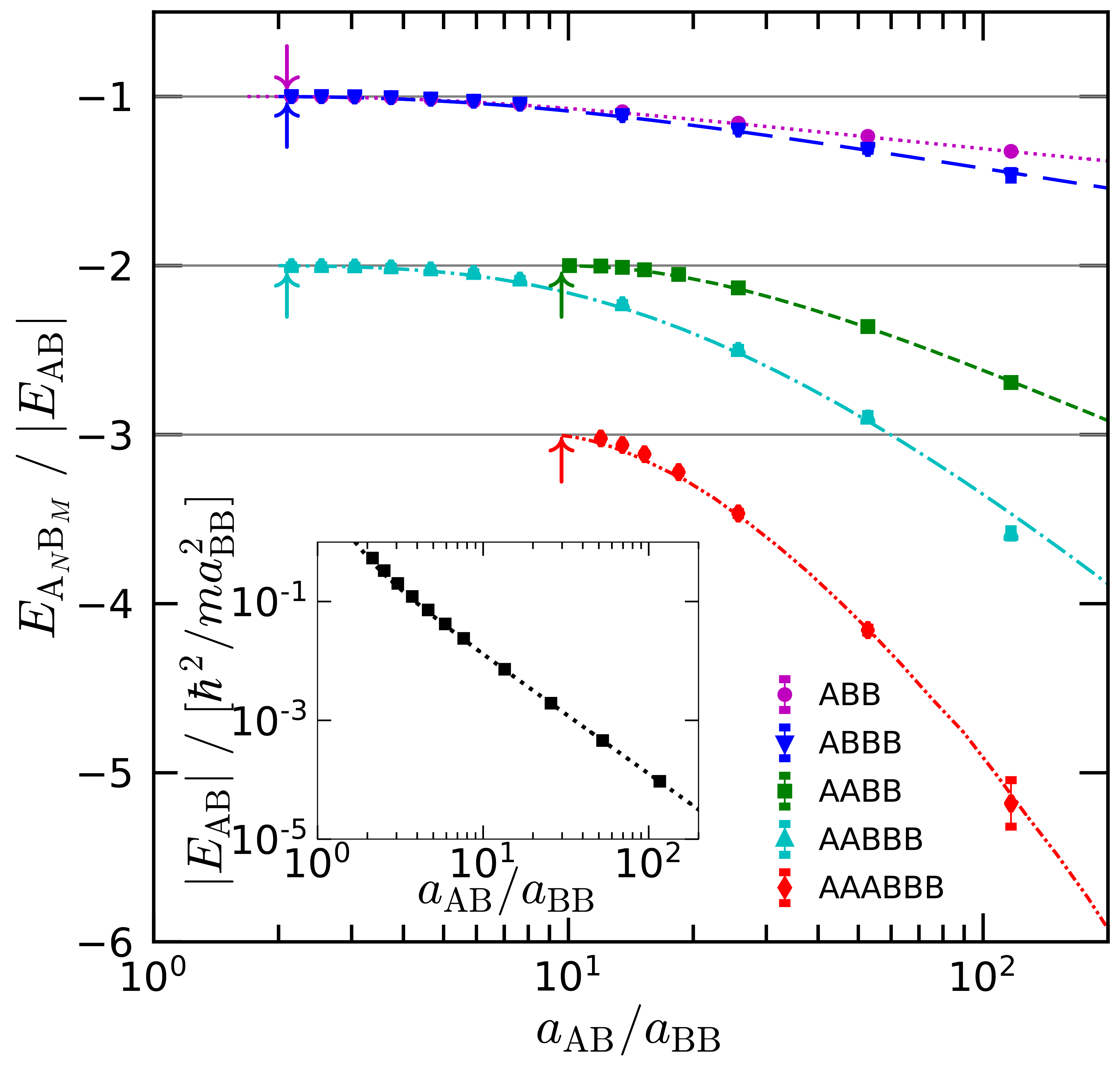}
	\caption{
	Energies of ${\rm A}_N{\rm B}_M$ clusters in units of the dimer energy $E_{\1\2}$ (reported in the inset in units of $\hbar^2/ma_{\2\2}^2$) for Gaussian (curves) and dipolar (symbols) potentials as a function of the scattering length ratio $a_{\1\2}/a_{\2\2}$. The arrows show the positions of the thresholds for binding in the dipolar case.
	}
  	\label{Fig:RatioEnergies}
\end{figure}

The Hamiltonian of our system is given by
\begin{equation}
\begin{aligned}
\label{Ham}
\hat H=&-\frac{\hbar^2}{2m}\sum_{i=1}^{N}\nabla^2_i-\frac{\hbar^2}{2m}\sum_{\alpha=1}^{M}\nabla_\alpha^2\\
+&\sum_{i<j}\hat V_{\1\1}(r_{ij})
+\sum_{\alpha<\beta} \hat V_{\2\2}(r_{\alpha\beta})
+\sum_{i,\alpha} \hat V_{\1\2}(r_{i\alpha}) \ ,
\end{aligned}
\end{equation}
where the two-dimensional vectors ${\bf r}_i$ and ${\bf r}_\alpha$ denote particle positions of species A and B containing, respectively, $N$ and $M$ atoms, $\hat V_{\1\2}$ and $\hat V_{\1\1}=\hat V_{\2\2}$ are the interspecies and intraspecies interaction potentials, and $m$ is the mass of each particle. For the bilayer setup, we have $V_{\1\1}(r) = V_{\2\2}(r) = d^2/r^3$ and $V_{\1\2}(r) = d^2(r^2-2h^2)/(r^2+h^2)^{5/2}$ where $d$ is the dipole moment and $h$ is the distance between the layers. Dipoles are aligned perpendicularly to the layers and there is no interlayer tunneling. The potential $V_{\2\2}(r)$ is purely repulsive and is characterized by the $h$-independent scattering length $a_{\2\2}=e^{2\gamma}r_0$ \cite{Ticknor2009}, where $\gamma\approx 0.577$ is the Euler constant and $r_0=md^2/\hbar^2$ is the dipolar length. 
The interlayer potential $V_{\1\2}(r)$ always supports at least one dimer state. 
Its energy reported in the inset of Fig.~\ref{Fig:RatioEnergies} diverges for $h\to 0$ and exponentially vanishes in the opposite limit~\cite{Shih2009,Armstrong2010,Klawunn2010,Baranov2011}. 
The scattering length $a_{\1\2}$, which is a function of $r_0$ and $h$, is $\sim a_{\2\2}\sim r_0$ for $h\sim r_0$, and exponentially large for $h \gg r_0$. In the following, we parametrize the system by specifying $a_{\2\2}$ and $a_{\1\2}$ rather than $h$ and $r_0$.

In the more academic case of Gaussian interactions, we use $\hat{V}_{\1\2}(r_{i\alpha})\psi(r_{i\alpha}) = \int V_{\1\2}(r_{i\alpha},r'_{i\alpha})\psi(r'_{i\alpha})d^2 r'_{i\alpha}$ and similarly for $V_{\1\1}$ and $V_{\2\2}$, where $V_{\sigma\sigma'}(r,r')=C_{\sigma\sigma'}G_\xi(r)G_\xi(r')$, $G_\xi(r)=(2\pi \xi^2)^{-1}\exp(-r^2/2\xi^2)$, and $\xi$ is the characteristic range of the potential. An advantage of this non-local potential is that the two-body problem can be solved analytically, giving $C_{\sigma\sigma'}^{-1}=\frac{m}{4\pi\hbar^2}\left[2\ln \frac {2\xi}{a_{\sigma\sigma'}}-\gamma\right]$. In the following, we vary the ratio $a_{\1\2}/a_{\2\2}$, with $a_{\2\2}=1.4\xi$ fixed. Note that the available ratio is limited to $a_{\1\2}/a_{\2\2}>1.1$.


In order to calculate the energies of the different few-body clusters, we employ two numerical techniques. In the dipolar case we use the diffusion Monte Carlo (DMC) method~\cite{BoronatCasulleras1994}, which leads to the exact ground-state energy of the system, within a statistical error. This stochastic technique solves the Schr\"odinger equation in imaginary time using a trial wave function for importance sampling. We choose it to be 
\begin{align*}\nonumber
\Psi_T(\mathbf{r}_1,\dots,\mathbf{r}_{N+M})&=\prod_{i<j}^{N}f_{\1\1}(r_{ij})\prod_{\alpha<\beta}^{M}f_{\2\2}(r_{\alpha\beta})\\
&\hspace{-0.5cm}\times\left[\prod_{i=1}^{N}\sum_{\alpha=1}^{M}f_{\1\2}(r_{i\alpha}) + \prod_{\alpha=1}^{M}\sum_{i=1}^{N}f_{\1\2}(r_{i\alpha})\right].
\end{align*}
The interspecies correlations are described by the dimer wave function $f_{\1\2}(r)$, calculated numerically, and the intraspecies Jastrow factors are chosen as the zero-energy two-body scattering solution, $f_{\1\1}(r)=f_{\2\2}(r)=K_0(2\sqrt{r_0/r})$, with $K_0$ the modified Bessel function.


In the Gaussian model, we use the stochastic variational method (SVM)~\cite{SuzVar98} where the wave function is expanded in a correlated Gaussian basis $\Psi(\evec)=\sum_i c_i \, \hat {\mathcal S} \,\exp\left({-}\frac{1}{2}\evec^T A_i \evec\right)$, where $\evec$ is the vector of $N+M-1$ particle coordinates in the center-of-mass reference frame, the matrices $A_i$ are real, symmetric, and positive definite. $\hat{\mathcal S}$ is the symmetrization operator, relevant for our Bose-Bose mixture, and the index $i$ sums over the basis functions (with $\approx 3000$ functions). The Schr\"odinger equation is then reduced to a generalized eigenvalue problem, giving the expansion coefficients $\{c_i\}$ and the corresponding energy. The basis is optimized to the system at hand in a stochastic way, where elements of the matrices $A_i$ are chosen randomly taking at each step the element that gives the lowest energy. Our SVM implementation closely follows Ref.~\cite{BazEliKol16}.


We first discuss the limit of very large $a_{\1\2}$ (large dimer size) when the interaction range and the intraspecies interactions can be neglected. In this case, the problem can be treated in the zero-range approximation giving for the ABB trimer $E_{\rm ABB}^{a_{\2\2}=0}=2.39 E_{\rm AB}$ \cite{Brodsky2006,PricoupenkoPedri2010,Bellotti2011} and for the tetramers $E_{\rm ABBB}^{a_{\2\2}=0}=4.1 E_{\rm AB}$ and $E_{\rm AABB}^{a_{\2\2}=0}=10.6 E_{\rm AB}$ \cite{Brodsky2006}. Here, we find that the other A$_N$B$_M$ clusters (with $1\leq N\leq M\leq 3$) are also bound in absence of the intraspecies repulsion. Using the method of Ref.~\cite{BazakPetrov2018}, we calculate their binding energies (and also update the energies of smaller clusters): 
$E_{\rm ABB}^{a_{\2\2}=0}/E_{\rm AB}=2.3896(1)$, 
$E_{\rm ABBB}^{a_{\2\2}=0}/E_{\rm AB}=4.1364(2)$, 
$E_{\rm AABB}^{a_{\2\2}=0}/E_{\rm AB}=10.690(2)$, 
$E_{\rm AABBB}^{a_{\2\2}=0}/E_{\rm AB}=28.282(5)$, and 
$E_{\rm AAABBB}^{a_{\2\2}=0}/E_{\rm AB}=104.01(5)$. 

The intraspecies repulsion shifts the cluster energies upwards as has been seen for the ABB trimer~\cite{PRB95_045401,BazakPetrovPRL2018} and for the ABBB tetramer~\cite{BazakPetrovPRL2018}. In Fig.~\ref{Fig:RatioEnergies}, we report the energies of these and bigger clusters for both the dipolar and Gaussian interactions. Note that, even for the weakest BB repulsion shown in this figure ($a_{\1\2}/a_{\2\2} = 200$), the clusters are significantly less bound compared to the case of no BB repulsion. This happens since the small parameter that controls the weakness of the intraspecies interaction relative to the interspecies one is actually $\lambda=1/\ln(a_{\1\2}/a_{\2\2}) \ll 1$. By contrast, effective-range corrections contain powers of $r_0\sqrt{mE}/\hbar$ or $\xi \sqrt{mE}/\hbar$ for dipolar or Gaussian interactions, respectively, which are exponentially small in terms of $\lambda$. This explains why the two interaction models lead to almost indistinguishable results for large $a_{\1\2}/a_{\2\2}$. 

\begin{figure}
	\centering
	\includegraphics[width=0.5\textwidth]{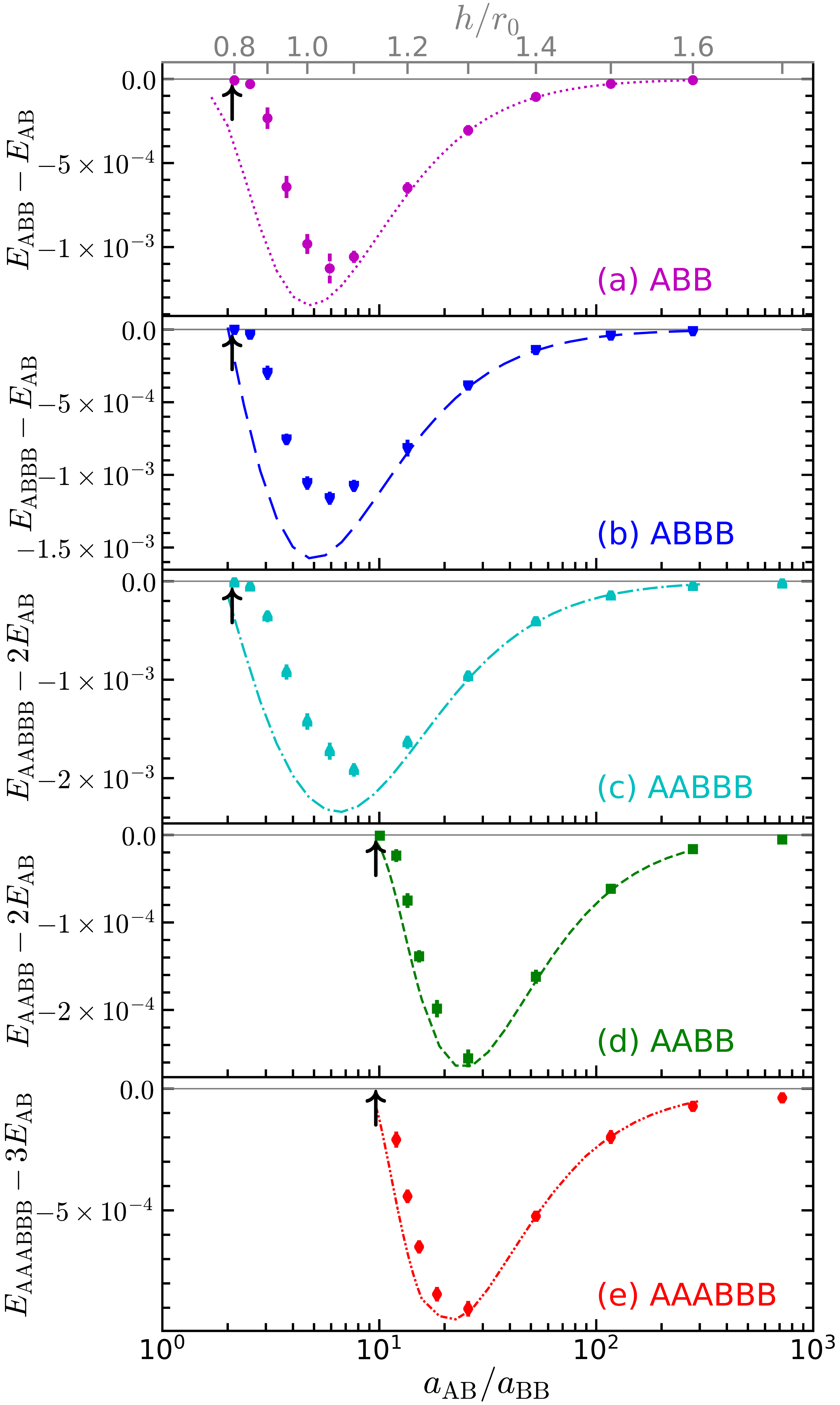}
	\caption{Binding energies of the few-body clusters $E_{{\rm A}_N{\rm B}_M}-NE_{\rm AB}$, in units of $\hbar^2/ma_{\2\2}^2$, versus $a_{\1\2}/a_{\2\2}$, for Gaussian (curves) and dipolar (symbols) potentials.}
	\label{BindingEnergies}
\end{figure}

We find that for sufficiently strong intraspecies repulsion (smaller $a_{\1\2}/a_{\2\2}$) the trimer and all higher clusters get unbound. In Fig.~\ref{Fig:RatioEnergies}, the thresholds for binding in the dipolar model are shown by arrows. We find that the tetramer threshold is located at $a_{\1\2}/a_{\2\2} \approx 10$ ($h/r_0\approx 1.1$) and the trimer threshold, corresponding to the atom-dimer zero crossing, occurs in the regime where all relevant length scales (scattering lengths, dimer sizes, interaction ranges) are comparable to one another; $a_{\1\2}/a_{\2\2} \approx 2$ ($h/r_0\approx 0.8$) for the dipolar model. The positions of the threshold and differences between the results of the two models are better visible in Fig.~\ref{BindingEnergies} where we plot the cluster energies in units of $\hbar^2/ma_{\2\2}^2$. The threshold values are obtained by fitting the energy results to the function $E_{{\rm A}_N{\rm B}_M}-NE_{\rm AB}=E_0\exp\{-1/[c_1(a_{\1\2}-a_{\1\2}^c) + c_2(a_{\1\2}-a_{\1\2}^c)^2]\}$, for $a_{\1\2}>a_{\1\2}^c$, where $a_{\1\2}^c,E_0,c_1,c_2$ are free parameters.

Our numerical calculations for larger clusters indicate that, depending on whether they are balanced ($M=N$) or not, their unbinding thresholds coincide, respectively, with the tetramer or with the trimer ones. To understand these results note that close to these thresholds the clusters are much larger than the dimer. Treating there the latter as an elementary boson D, the AABBB pentamer and the ABBB tetramer can be thought of as weakly bound DDB or DBB ``trimers'' characterized by a large $a_{{\rm D}{\rm B}}$ value and repulsive DD and BB interactions (the DD interaction is repulsive since we are above the tetramer AABB threshold). In the limit $a_{{\rm D}{\rm B}}\rightarrow \infty$ the DD and BB interactions can be neglected and the binding energies of the DDB and DBB composite trimers are asymptotically fractions of $E_{\1\2\2}-E_{\1\2}$~\cite{PricoupenkoPedri2010}. The ABB trimer, ABBB tetramer, and AABBB pentamer thresholds are therefore the same [see Fig.~\ref{BindingEnergies}~(a,b,c)]. In the same reasoning, close to the AABB tetramer crossing, the hexamer AAABBB is a weakly-bound DDD state which splits into three dimers when the dimer-dimer attraction changes to repulsion resulting in the same threshold value.

In the above discussion we have integrated out the internal degrees of freedom of the dimers, replacing them by elementary point-like bosons. In fact, the DD zero crossing that we observe for $a_{\1\2}\approx 10 a_{\2\2}$ is a nonperturbative phenomenon resulting from a competition between strong repulsive and attractive interatomic forces among four individual atoms. These interactions are strong since the corresponding scattering lengths are comparable to the typical atomic de Broglie wave lengths $\sim 1/a_{\rm AB}$. We emphasize that this cancellation is achieved only for two dimers. For three dimers it is incomplete and there is a residual effective three-dimer force of range $\sim a_{\rm AB}$ (distance, where the dimers start touching one another). In the many-body problem, this higher-order force may compete with the dimer-dimer interaction (if it is not completely zero) or even become dominant. In principle, one can also discuss higher-order effects of this type at the DB zero crossing in a DB mixture, but they are expected to be subleading since the DD and BB interactions remain finite. In the remainder of this Rapid Communication we thus concentrate on the population-balanced case.

\begin{figure}
	\centering
	\includegraphics[width=0.45\textwidth]{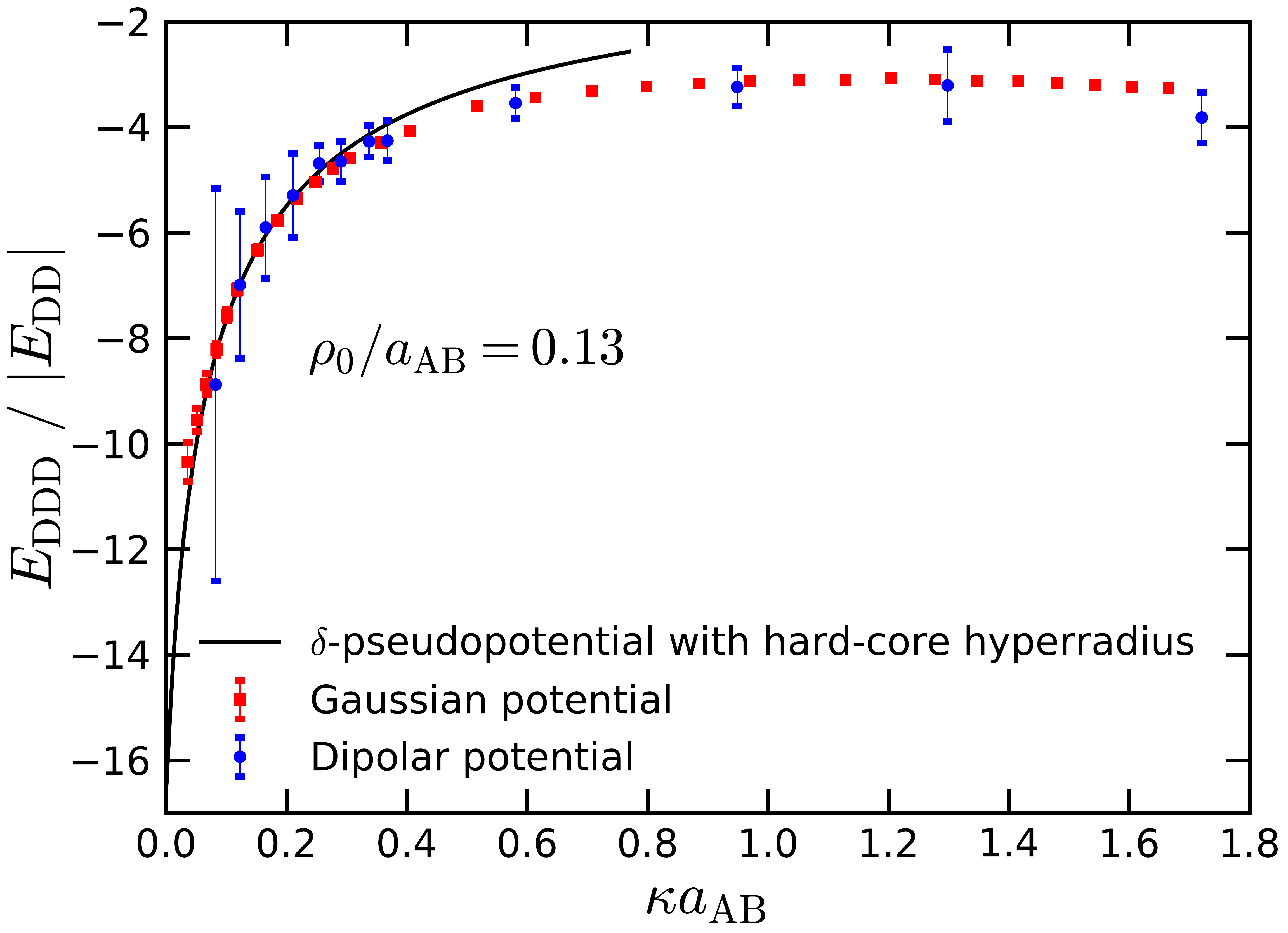}
	\caption{The hexamer-to-tetramer binding energy ratio $E_{\rm DDD}/|E_{\rm DD}|$ as a function of $\kappa a_{\rm AB}$. The solid line is the result of the zero-range model with the hard-core hyperradius constraint at $\rho_0=0.13a_{\rm AB}$.}
	\label{Fig:ThreeBodyForce}
\end{figure}

In order to characterize the effective three-dimer interaction, we follow the method developed previously in one dimension~\cite{Guijarro2018}. Namely, we analyze the behavior of the hexamer energy just below the tetramer threshold. If the tetramer binding energy $E_{\rm DD}=E_{\rm AABB}-2E_{\rm AB}$ is much smaller than $E_{\1\2}$, the dimer-dimer interaction can be considered point-like and the relative DD wave function can be approximated by $\phi(r)\propto K_0(\kappa r)$, where $\kappa=\sqrt{-2mE_{\rm DD}/\hbar^2}$ is the inverse size of the tetramer. Similarly, the AAABBB hexamer under these conditions reduces to the well-studied problem of three point-like bosons~\cite{BruchTjon1979,Adhikari1988,Nielsen1997,Nielsen1999,HammerSon2004,KartavtsevMalykh2006,Brodsky2006}, according to which the ground state hexamer binding energy $E_{\rm DDD}=E_{\rm AAABBB}-3E_{\rm AB}$ should satisfy~\cite{HammerSon2004,KartavtsevMalykh2006}
\begin{equation}\label{E3zr}
E_{\rm DDD}/|E_{\rm DD}|=-16.5226874.
\end{equation}
We expect the ratio $E_{\rm DDD}/|E_{\rm DD}|$ to reach the zero-range limit~(\ref{E3zr}) as we approach the dimer-dimer zero crossing, i.e., as $\kappa a_{\rm AB}\rightarrow 0$. In Fig.~\ref{Fig:ThreeBodyForce}, we plot $E_{\rm DDD}/|E_{\rm DD}|$ versus $\kappa a_{\rm AB}$ and indeed see a tendency towards the value~(\ref{E3zr}) although the effects of the finite size of the dimers and their internal degrees of freedom, that we have neglected in the zero-range model, are obviously important. The fact that the hexamer energy lies above the limit~(\ref{E3zr}) points to an effective three-dimer repulsive force. We note again that the values of the ratio $E_{\rm DDD}/|E_{\rm DD}|$ obtained for Gaussian and dipolar potentials are quite close to each other for all values of $a_{\1\2}$ suggesting a certain universality of this problem and a relative unimportance of the long-range interaction tails.

In order to quantify the three-dimer interaction observed in Fig.~\ref{Fig:ThreeBodyForce}, we extend the model of three point-like dimers by requiring that the three-dimer wave function vanishes at a hyperradius $\rho_0$. For three dimers, with coordinates ${\bf r}_1$, ${\bf r}_2$, and ${\bf r}_3$, the hyperradius is defined as $\rho=\sqrt{x^2+y^2}$, where ${\bf x}=(2{\bf r}_3-{\bf r}_1-{\bf r}_2)/\sqrt{3}$ and ${\bf y}={\bf r}_1-{\bf r}_2$ are the Jacobi coordinates. Clearly, for this minimalistic model $E_{\rm DDD}/|E_{\rm DD}|$ is a function of the ratio $\kappa \rho_0$, relating the three- and two-dimer interaction strengths. Given the isotropic form of the three-body constraint in the hyperradial space, a natural way of solving this problem is to use the adiabatic hyperspherical method. Kartavtsev and Malykh~\cite{KartavtsevMalykh2006} discussed this method in detail and applied it to the $\rho_0=0$ limit of our problem, i.e., the case of no three-body interaction. The only modification of their procedure, to account for finite $\rho_0$, is to set the hyperradial channel functions to zero at $\rho=\rho_0$. In this way, we obtain the ratio $E_{\rm DDD}/|E_{\rm DD}|$ as a function of $\kappa \rho_0$. We then treat $\rho_0$ as a constant (independent of $E_{\rm DD}$) determined by fitting the DMC and SVM data in $\kappa a_{\rm AB} < 0.4$ range. By minimizing $\chi^2$ we obtain $\rho_0=0.13 a_{\rm AB}$.

The inclusion of the three-body hard-core constraint, even corresponding to numerically very small $\kappa\rho_0$, leads to a spectacular deviation from Eq.~(\ref{E3zr}). This interesting effect is due to an enhancement of the three-dimer interaction by strong two-dimer correlations. Indeed, in the absence of two-body interactions, the three-body scattering in two dimensions is equivalent to the four-dimensional two-body scattering on a short-range potential. The corresponding hyperradial wave function for $\rho$ larger that the support of the potential, but smaller than the de Broglie wave length, is proportional to $1-\rho_0^2/\rho^2$. The same scattering effect in the first Born approximation (and the same mean-field interaction shift) is attained by using the three-body potential $V_3({\bf r}_1,{\bf r}_2,{\bf r}_3)=3\pi^2(\hbar^2\rho_0^2/m) \delta({\bf r}_1-{\bf r}_2)\delta({\bf r}_2-{\bf r}_3)$. Naively, we expect the leading small-$\rho_0$ correction to Eq.~(\ref{E3zr}) to behave as $(\kappa \rho_0)^2$ as if the three dimers without two-body interactions were externally confined to a surface $\sim 1/\kappa ^2$. However, in our case two-dimer correlations are strong and the three-dimer wave function is logarithmically enhanced at short hyperradii~\cite{KartavtsevMalykh2006}. More precisely, by using arguments of Ref.~\cite{KartavtsevMalykh2006} one can show that the hyperradial wave function at distances $\rho \ll 1/\kappa$ behaves as $\ln^3(\kappa\rho)-\rho_0^2\ln^6(\kappa \rho_0)\rho^{-2}\ln^{-3}(\kappa \rho)$ and the leading-order correction to Eq.~(\ref{E3zr}) behaves as $\sim(\kappa \rho_0)^2\ln^6(\kappa \rho_0)$, representing a noticeable enhancement compared to the case of no two-body interaction.

Promising candidates for observing the predicted cluster states are bosonic dipolar molecules characterized by large and tunable dipolar lengths which, at large electric fields, tend to $r_0=5\times 10^{-6}$m for $^{87}$Rb$^{133}$Cs~\cite{PhysRevLett.113.205301,PhysRevLett.113.255301}, $r_0 = 2\times10^{-5}$m for $^{23}$Na$^{87}$Rb~\cite{PhysRevLett.116.205303,PhysRevA.97.020501} and $r_0=6\times10^{-5}$m for $^7$Li$^{133}$Cs~\cite{Deiglmayr2010}. Fermionic $^{87}$Rb$^{40}$K~\cite{Moses2015,DeMarco2019} and $^{23}$Na$^{40}$K~\cite{Park2015,Park2017,Yang2019} molecules ($r_0=7\times 10^{-7}$m and $r_0=7\times 10^{-6}$, respectively) could be turned into bosons by choosing another isotope of K. The interlayer distance, fixed by the laser wavelength, has typical values of $h\approx (2-5)\times 10^{-7}$m, which is thus sufficient for observing the few-body bound states that we predict for ratios $h/r_0 > 0.8$. 

A subject of further work is to generalize these findings to the many-body problem when a new scale (density $n$) comes into play. It is important to understand how the two- and three-body effects correlate with each other as one passes through the dimer-dimer zero crossing. Although we find no qualitative difference between the dipolar and Gaussian models in our few-body results, the long-range tails will be important when the quantity $nr_0$ becomes comparable to the inverse healing length (which is where the dipolar condensate becomes rotonized). For bilayer dipoles the relevant region of parameters is close to the dimer-dimer zero crossing, which we predict to be at $h/r_0\approx 1.1$. Finally, it is instructive to consider a simpler model of elementary bosons with (possibly exotic) finite-range interactions and investigate whether finite-range effects can in general be absorbed into an effective three-body interaction. Systematic calculations of the trimer-to-dimer binding energy ratio (see \cite{Adhikari1988}) could then serve as a tool for characterizing this effective force.

To summarize, we have studied few-body clusters A$_N$B$_M$ with $N\leq M\leq 3$ in a two-dimensional Bose-Bose mixture using different (long-range dipolar and short-range Gaussian) intraspecies repulsion and interspecies attraction models. In both cases, the intraspecies scattering length $a_{\1\1}=a_{\2\2}$ is of the order of the potential ranges, whereas we tune $a_{\1\2}$ by adjusting the AB attractive potential (or the interlayer distance in the bilayer setup). We find that for $a_{\1\2}\gg a_{\2\2}$ all considered clusters are (weakly) bound and their energies are independent of the interaction model. As the ratio $a_{\1\2}/a_{\2\2}$ decreases, the increasing intraspecies repulsion pushes the clusters upwards in energy and eventually breaks them up into $N$ dimers and $M-N$ free B atoms. In the population balanced case ($N=M$) this happens at $a_{\1\2}/a_{\2\2}\approx 10$ where the dimer-dimer attraction changes to repulsion. By studying the AAABBB hexamer near the dimer-dimer zero crossing we find that it very much behaves like a system of three point-like particles (dimers) characterized by an effective three-dimer repulsion. A dipolar system in a bilayer geometry can thus exhibit the tunability and mechanical stability necessary for observing dilute liquids and supersolid phases.

This study has been partially supported by funding from the Spanish MINECO (FIS2017-84114-C2-1-P). The Barcelona Supercomputing Center (The Spanish National Supercomputing Center - Centro Nacional de Supercomputaci\'on) is acknowledged for the provided computational facilities (RES-FI-2019-2-0033). G.G. acknowledges a fellowship by CONACYT (M\'exico). D.S.P. acknowledges support from ANR Grant Droplets No. ANR-19-CE30-0003-02.


%

\end{document}